\documentclass[aps,prl,twocolumn,groupedaddress,showpacs]{revtex4}

\usepackage{textcomp}
\usepackage{graphicx}
\usepackage{braket}
\usepackage{subfigure}
\usepackage{psfrag}

\begin{document}

\title{Atomic Landau-Zener tunneling in Fourier-synthesized optical lattices}

\author{Tobias Salger}
\author{Carsten Geckeler}
\author{Sebastian Kling}
\author{Martin Weitz}
\affiliation{Institut f\"ur Angewandte Physik, Universit\"at Bonn,
  Wegelerstr. 8, 53115 Bonn}

\date{\today}

\renewcommand{\figurename}{Fig.}
\begin{abstract}
We report on an experimental study of quantum transport of atoms in
variable periodic optical potentials. The band structure of both
ratchet-type asymmetric and symmetric lattice potentials is
explored. The variable atom potential is realized by superimposing a
conventional standing wave potential of $\lambda/2$ spatial
periodicity with a fourth-order multiphoton potential of $\lambda /4$
periodicity. We find that the Landau-Zener tunneling rate between the
first and the second excited Bloch band depends critically on the
relative phase between the two spatial lattice harmonics.
\end{abstract}

\pacs{03.75.Lm, 32.80.Pj, 42.50.Vk}

\maketitle

Transport properties of quantum objects subject to a periodic potential
are determined by the particle's band structure. The
energy spectrum here splits into continous
energy bands separated by bandgaps. For example, the more than 20 orders of
magnitude difference in electrical conductivity between an isolator and a
good conductor thus finds a natural physical explanation~\cite{kittel}. In recent
years, atoms confined in periodic optical potentials, so called optical
lattices, have developed a powerful tool for the observation of effects
known or predicted in solid state physics~\cite{Bloch-lattice}. So far, the band structure
has  been exploited only for sinusoidal lattice potentials, as can be
realized with the ac Stark shift of optical standing waves. In remarkable
experiments with such standing wave lattices, Bloch oscillations and
Landau-Zener transitions have been
observed~\cite{Salomon-Bloch,Raizen-Wannier,Raizen-Landau}.

Here we report on experiments studying the band structure of optical
lattices with variable inversion symmetry and shape, as a step towards
simulating the variety of potential forms that nature provides us in the
system of electrons in natural crystals. The used potentials are realized by
superimposing a conventional standing wave lattice of $\lambda /2$ spatial
periodicity with a $\lambda /4$ periodicity lattice realized using the
dispersion of higher order Raman transitions. By varying the phase between
the two spatial harmonics, symmetric and ratchet-type asymmetric lattice
potentials are realized, which exhibit a different band structure. We
experimentally demonstrate that the strength of interband transitions for an
atomic Bose-Einstein condensate depends on the shape of the lattice
potential.

Before proceeding, let us point out that in semiconductor
heterostructures effects of the inversion symmetry of quantum wells have
been studied using magnetotransport~\cite{Eisenstein-heterostructure}. In the area of atomic physics,
directed transport has been achieved in driven ratchet systems with the
temporal symmetry broken by dissipative processes~\cite{Renzoni-dissi}. Further, in
theoretical works, atom transport has been studied in periodic double well
systems~\cite{Korsch-Zener}.
Let us begin by describing our calculations of the band structure in a
Fourier-synthesized atom potential realized by superimposing two
lattice potentials of spatial periodicities $\lambda/2$ and $\lambda/4$:
\begin{equation}
V(z)=\frac{V_1}{2}\cos (2kz)+\frac{V_2}{2}\cos (4kz+\varphi)
\end{equation}
where V$_1$ and V$_2$ denote the potential depths of the two lattice
harmonics respectively and $\varphi $ the relative phase. According to
Bloch's theory~\cite{Bloch-Theory}, the band structure of the periodic potential can be
derived by
solving the eigenvalue equation $Mc_{l}^{q}=E_{q}^{(n)} c_{l}^{q},$ where the
quasimomentum q conventionally is restricted to the first Brillouin zone: $%
-\hbar k< q<\hbar k.$ Here we search for the Eigenenergies
E$_{q}^{(n)}$ of the Eigenstates $\ket{n,q}=\sum_l c_{l}^{q}e^{i2lkz}$ using the coupling matrix M
with elements $M_{jj}=(2l\hbar k+q)^{2}/2m,
M_{j,j+1}=M_{j+1,j}^{\ast}=V_{1}/4,$ and
$M_{j,j+2}=M_{j+2,j}^{\ast}=V_{2}/4\cdot e^{i\varphi },$ where the
index n denotes the band 
number. The matrix can be readily diagonalized. For the
lattice potential of eq. (1), Fig.~1 shows a spatial lattice potential
(left) and the corresponding band structure (right) for different
values of the relative phase $\varphi $ of the
two spatial lattice harmonics. It is clearly visible that the gap between
the first and second excited band is strongly dependent on the value of the
relative phase, while no such significant modification of the splitting between the
other shown bands is visible. Physically, the variation of this splitting on
the relative phase of the lattice harmonics can be understood by the
interference of the second order Bragg scattering amplitude of the standing
wave potential of periodicity $\lambda /2$ with the first order Bragg
scattering amplitude of the $\lambda /4$ periodicity multiphoton lattice
potential, which both contribute to this bandgap. In the limit of a very
shallow lattice with the four-photon contribution being a small
perturbation, i.e. V$_{2}\ll V_{1}\ll E_{r}$ (where $E_r=\hbar^2k^2/2m$
denotes the atomic recoil energy), the size of this bandgap is
determined by the simple analytical expression
$|V_{1}^{2}/16E_{r}+e^{i\varphi }\cdot V_{2}|/2$, which directly shows
the two interfering contributions of coupling Rabi frequencies arising
from different lattice harmonics. For
larger potential values, higher order corrections come into play, but
by numerical diagonalization of the coupling matrix M, the band structure
for arbitrary potential values is readily determined. The bandgap
reaches a maximum
value for $\varphi =0^\circ$ in which case the lattice potential
resembles a periodic sequence of
hills (Fig.~1a). On the other
hand, the bandgap reaches its minimum value for $\varphi=180^\circ,$ corresponding
to an array of potential dimples in the spatial lattice structure (Fig.~1c).
For a suitable choice of potential values, the bandgap
between first and second excited band can even disappear.
The situation of spatial lattice potentials with broken spatial symmetry,
as, e.g. the saw-tooth like structures shown in Fig.~1b for $\varphi
=\pm 90^\circ,$ yields an intermediate value of the bandgap.
\begin{figure}
\includegraphics[width=0.9\linewidth]{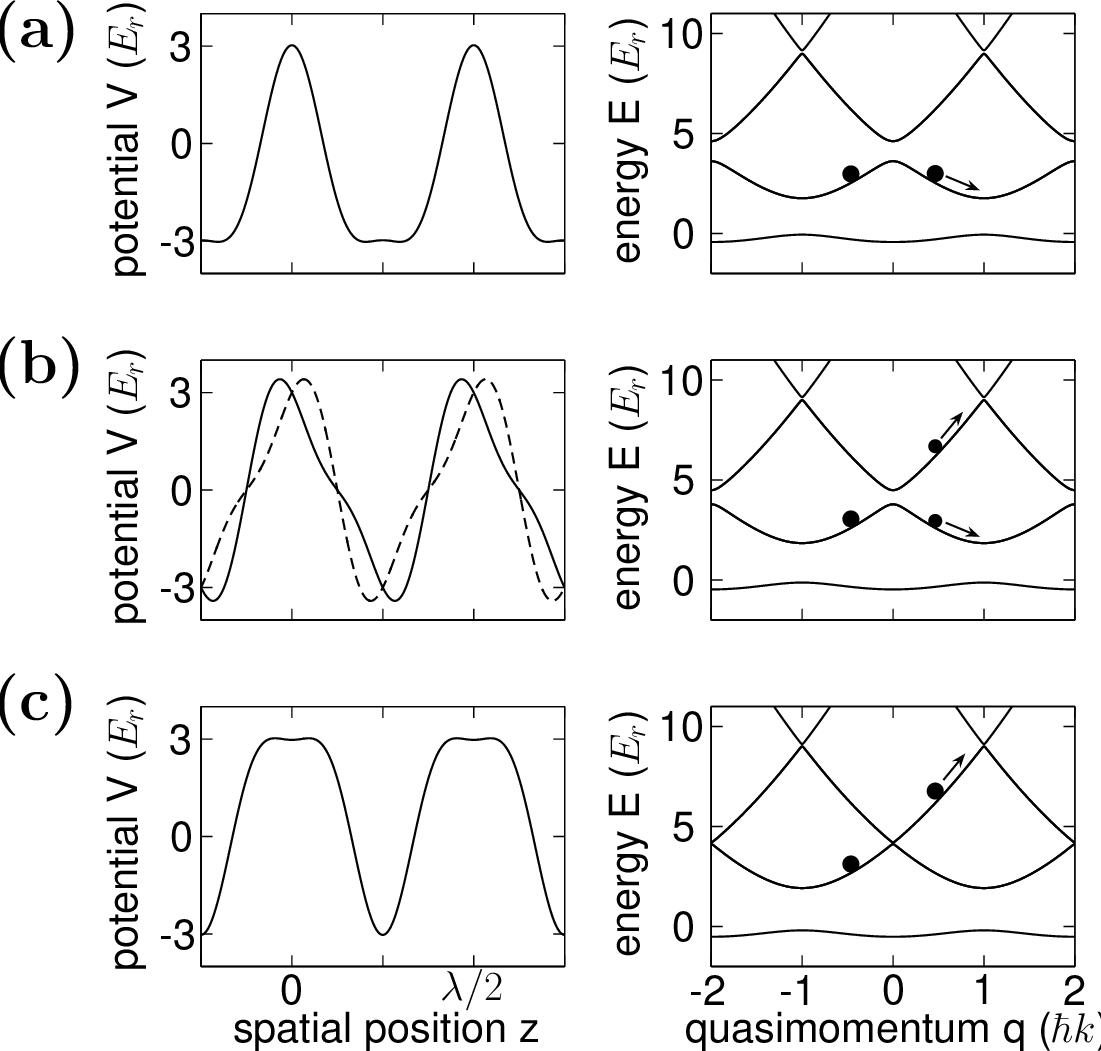}%
\caption{Spatial potential (left) and band structure
   (right) for a periodic atom potential $V(z)=V_1\cos
    (2kz)/2+V_2\cos(4kz+\varphi)/2$ for different values of the phase
    $\varphi$
    between lattice harmonics: (a) $\varphi=0^\circ$: periodic
    sequence of hills, (b) $\varphi=90^\circ$ (solid line) and
    $\varphi=-90^\circ$ (dashed line): sawtooth-like potentials, (c)
    $\varphi=180^\circ$: periodic sequence of dimples. Here,
    V$_{1}=4 E_{r}$ and V$_{2}=1.2 E_{r}$ were used for the sake of
    clarity as an example for a set of potential values in which the
    splitting between first and second excited Bloch band vanishes at
    $\varphi=180^\circ$. The size of this bandgap can be studied
    by Landau-Zener tunneling of atoms, as indicated in the
    plots.}
\end{figure}
We experimentally exploit the band structure of the Fourier-synthesized
lattice by means of quantum transport experiments. Specifically, the size of
the gap between first and second excited Bloch bands is measured by means of
Landau-Zener tunneling of atoms in an accelerated lattice
potential. The 
acceleration provides an inertial force in the moving lattice frame,
emulating a force on atomic wave packets. The Landau-Zener tunneling
probability can
be estimated to be $\Gamma =\exp (-a_{c}/a),$ where $a_{c}=\pi\Delta
^{2}/(8\hbar^2 k)$ with $\Delta $ denoting the width of the energy
gap.

Our method for generating a lattice potential with variable spatial
symmetry and form is similar as described previously~\cite{Ritt-Fourier}. For the generation of
the fundamental frequency, we use a conventional standing wave
lattice potential, as achieved with two counterpropagating optical waves
with frequency $\omega $ detuned from an atomic resonance. The resulting
potential $V(z)=-(\alpha /2)|E(z)|^{2}$, where $\alpha $ denotes the dynamic
atomic polarizability and E(z) the electric field, is proportional to $%
\cos ^{2}kz=(1+\cos 2kz)/2,$ yielding the well known $\lambda /2$ spatial
periodicity of optical standing waves. In a quantum picture, the atoms
undergo virtual two-photon processes of absorption of a photon from one mode
followed by simulated emission into the counterpropagating mode. In
principle, a potential with periodicity $\lambda /4$
could be achieved by replacing the absorption
and the stimulated emission cycle by a four-photon process induced by
photons of wavelength $\lambda $, as indicated in Fig.~2a (right). The spatial
periodicity of the achieved multiphoton lattice contribution is $\lambda
_{eff}/2=\lambda /4,$ where $\lambda _{eff}=\lambda /2$ denotes the
effective wavelength of a two-photon field. Fig.~2b shows the used scheme
for a multiphoton lattice potential with periodicity $\lambda /4,$ which is
based on a three-level configuration with two stable ground states 
$\ket{g_0}$ and $\ket{g_1}$ and
one spontaneously decaying excited state
$\ket{e}$~\cite{Cohen-multi,Weitz-multi}. Compared to the four-photon
ladder scheme, in this improved
approach one absorption (stimulated emission) process has been replaced by a
stimulated emission (absorption) process of a counterpropagating photon
respectively. A minimum of three laser frequencies is required to suppress
standing wave effects, and the atoms are irradiated with two optical beams
of frequencies $\omega +\Delta \omega $ and $\omega -\Delta \omega $ from
one side and a further beam of frequency $\omega $ from the
counterpropagating direction. The high frequency resolution of Raman
spectroscopy here allows to clearly separate in frequency space the desired
four-photon process from lower order contributions. The described scheme can
be extended to higher lattice periodicities, where in general an effective
potential with periodicity $\lambda /2n$ can be achieved by a 2n-th order
process~\cite{Weitz-multi}. By combining lattice potentials of different spatial
periodicities, arbitrarily shaped periodic potentials can be synthesized.
\begin{figure}
\includegraphics[width=0.9\linewidth]{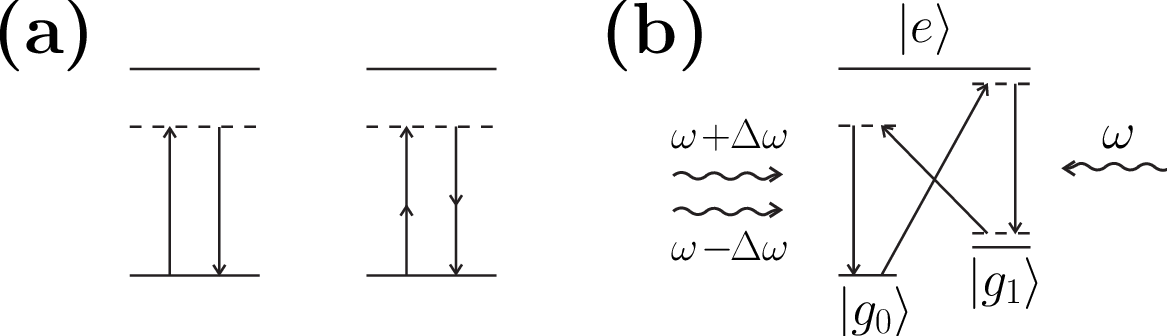}%
\caption{(a) Left: Virtual two-photon process in a conventional
  standing wave lattice with $\lambda /2$ spatial
  periodicity. Right: Virtual process
  contributing to a lattice potential with $\lambda_{eff}/2=\lambda
  /4$ periodicity in
  a ladder level scheme. However, unwanted standing wave effects
  dominate in this simple approach. (b) Improved configuration for
  realization of a four-photon lattice with $\lambda /4$ spatial
  periodicity. This scheme is used in our experiments to generate a
  second spatial lattice harmonic.}
\end{figure}
Our experimental setup has been described
in~\cite{Ritt-Fourier,Cennini-BEC}. Briefly, light for
generation of variable atomic lattice potentials is produced by a tapered
diode laser tuned some 2 nm to the red of the rubidium D2-line. The emitted
radiation is split into two, and each of the partial beams pass an
acoustooptic modulator. The modulators are used for beam switching and
to superimpose several optical frequencies onto a single beam path, as is
required to generate superpositions of a standing wave potential and a
four-photon lattice potential by realizing the scheme of
Fig.~2b. After passing the modulators, the two beams are directed
through optical
fibers and send in a counterpropagating geometry onto a rubidium ($^{87}$Rb)
Bose-Einstein condensate.

Our Bose-Einstein condensate is produced all-optically by evaporative
cooling of $^{87}$Rb atoms in a CO$_{2}-$laser dipole trap. During the
evaporation, a magnetic field gradient is activated, resulting a
spin-polarized condensate with 1.6 x 10$^4$ atoms in the $\ket{F=1, m_F=-1}$
ground state. A magnetic bias field generates a frequency
splitting of $\omega _{z}\simeq 2\pi \cdot 805kHz$ between neighbouring
Zeeman ground states. The direction of the magnetic field forms an angle
respectively to the optical beam, so that atoms experience $\sigma ^{+}$-$%
,\sigma ^{-}$- and $\pi $-polarized light simultaneously. For generation of
a multiphoton lattice potential with the scheme of Fig.~2b, the F = 1 ground
state components m$_F$ = -1 and 0 are used as states $\ket{g_0}$ and $\ket{g_1}$, while the 
5P$_{3/2}$-manifold serves as the excited state 
$\ket{e}$. The Raman detuning $\delta $ is
2$\pi \cdot 50kHz.$ The used potential depths are V$_{1}\simeq
3 E_{r}$ and V$_{2}\simeq 2 E_{r}$ for the lattice contributions with periodicities 
$\lambda /2$ and $\lambda /4$ respectively, and
different values of the phase $\varphi $ between the two spatial harmonics
were used in the course of the experiments. Experimentally, the potential
values of both lattice harmonics and the phase $\varphi $ can be monitored
by a series of Raman-Nath diffraction experiments on pulsed optical
potentials and Rabi-oscillations~\cite{Mellish-ground,Ritt-Fourier}, so that all
parameters of the Fourier synthesized lattice potential of eq. 1 are known.
One of the lattice beams with frequency $\omega $ is used for
both the standing wave and the four-photon lattice potential. When this beam
is acoustooptically detuned by a small amount \ $\delta _{Dopp},$ the
reference frame in which the optical potential is stationary moves with a
velocity v$_{rel}=\delta _{Dopp}/2\pi \cdot \lambda /2,$ where $\lambda $
denotes the laser wavelength. We adiabatically load the atomic Bose-Einstein
condensate into the first band by transfering the atoms into a lattice
potential moving with $v_{rel}\equiv q_0/m\simeq 1.5\ $ $\hbar k/m.$

The lattice beams form an angle of 41$^\circ$
relatively to the axis of gravity, and the ballistic free
atomic fall
accelerates the atoms over the bandgap between the first and second excited
Bloch band. Fig.~3 shows the result of a measurement monitoring for
different final values of the atomic
quasimomentum. Here, two different lattice forms were investigated.
For a phase shift $\varphi \simeq 0^\circ$ (dots),
corresponding to a potential form with a sequence of hills, atoms
are Bragg-diffracted at the bandgap towards higher velocities. In
contrast, for a phase shift
$\varphi \simeq 180^\circ$ (crosses), corresponding
to a lattice consisting of a periodic sequence of dimples,
the ballistic free atom flight is hardly modified by a bandgap between the
first two excited bands. We attribute this striking modification of
transport properties on the potential shape to the strong dependence of the
size of the bandgap between the first two excited Bloch bands on the 
relative phase $\varphi$ between lattice harmonics. For a phase shift $\varphi$
=180$^\circ,$ almost all atoms undergo Landau-Zener transitions over
the bandgap near q = 2$\hbar k$ (corresponding to the second gap at q
= 0 in the reduced zone scheme), while adiabaticity is better achieved
with $\varphi =0^\circ,$ giving evidence for an increased splitting
of the bandgap. At the bandgap, Bragg diffraction
changes the atomic momentum in units of 4$\hbar k.$
\begin{figure}
\includegraphics[width=0.9\linewidth]{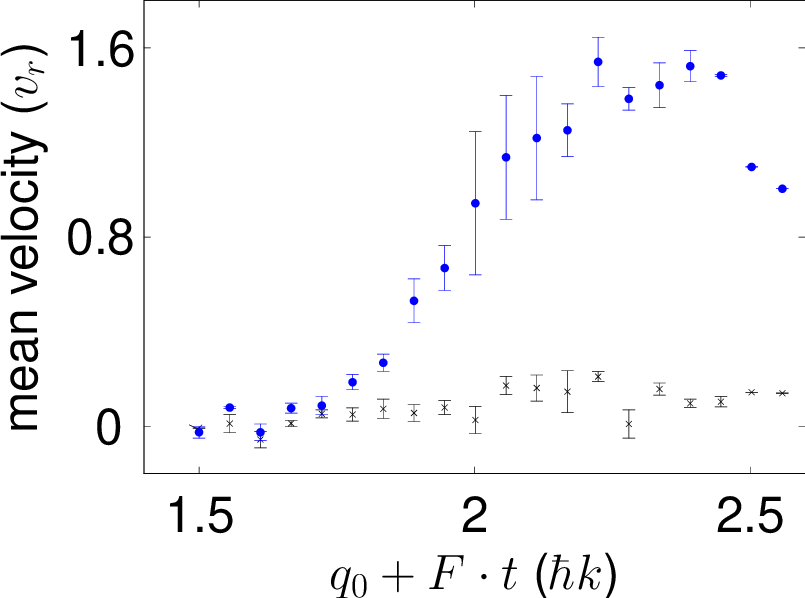}%
\caption{Temporal variation of the mean atomic velocity in units of
  the recoil velocity ($v_r=\hbar k/m$) for an atom
  with initial momentum of $q_0$=1.5 $\hbar k$ subject to the earth's
  gravitational field and the lattice potential for a phase $\varphi$
  of 0$^\circ$ (dots) and 180$^\circ$ (crosses) between lattice
  harmonics. The earth's gravitational force along the lattice
  axis is $F=m\cdot g\cdot \cos\alpha$, where $\alpha \simeq
  41^\circ$.}
\end{figure}
For a more detailed investigation of the bandgap
we have recorded the Landau-Zener tunneling rate as a
function of the phase between the lattice harmonics. For this measurement we
have increased the beam detuning $\delta _{Dopp}$ with a constant rate, so
that the lattice potential is accelerated relatively to the atomic frame
with an acceleration of 6.44~m/s$^{2}$, somewhat exceeding the
projection of the earth's gravitational field onto the beam
axis. Fig.~4
shows experimental data for the fraction of tunneled atoms
as a function of phase $\varphi$ between the two
lattice harmonics. The data fits well to a sinusoidal curve, as shown by the
solid line. Notably, asymmetrically shaped ratchet-like potentials result in
an intermediate value of the Landau-Zener tunneling rate, while smallest
(largest) values are achieved for hill-type (dimple-type) periodic
arrays. It is interesting to note that this characteristics is in contrast
to the behaviour in dissipative lattices, where maxima and minima of the
particle transport are achieved for ratchet-like potentials of different
symmetry. Experimentally, a related observation has been made in pulsed,
driven ratchet potentials~\cite{Renzoni-dissi}.
\begin{figure}
\includegraphics[width=0.9\linewidth]{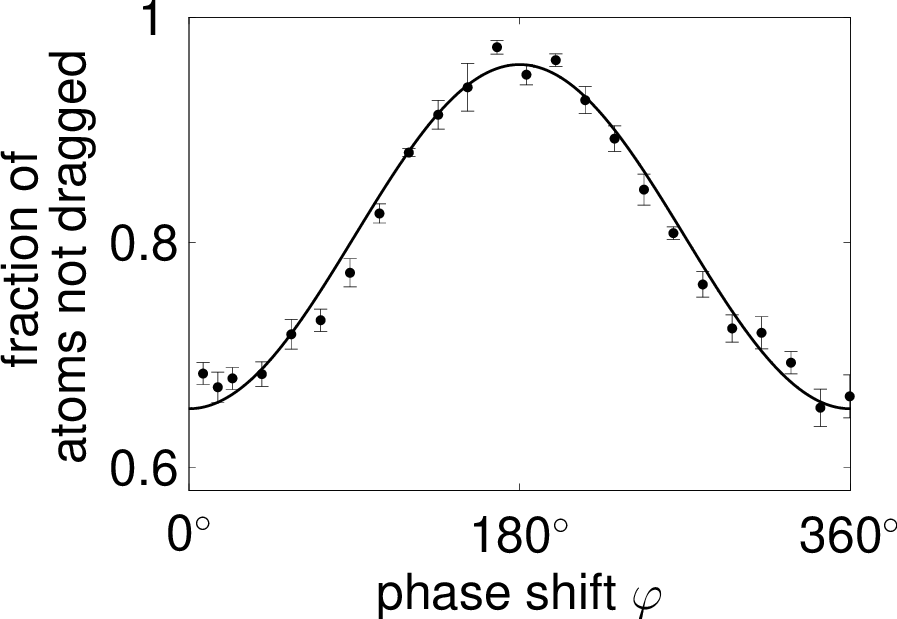}%
\caption{Fraction of atoms that have tunneled through the energy gap
  between first and second excited Bloch bands as a function of phase
  $\varphi$ between spatial lattice harmonics. The experimental data
  (dots) has been fitted with a sinusoidal curve (solid line).}
\end{figure}
In subsequent experiments, we have studied the variation of the Landau-Zener
tunneling rate on the depth of the optical standing wave contribution to the
total optical potential. The four-photon contribution with periodicity of $%
\lambda /4$ here was left constant (see eq. 1). Fig.~5 shows experimental
data for the interband tunneling for a phase shift $\varphi
=0^\circ$ (dots) and $\varphi =180^\circ$ (crosses). For the former
phase shift value, the tunneling rate decreases with the standing wave
contribution V$_{1}$ for all parts of the curve, as is consistent with
a monotonely increasing energy gap between the bands. On the other
hand, for a phase shift $\varphi =180^\circ$ between lattice harmonics
a local maximum of the tunnelling rate is
observed for an intermediate value of V$_{1}$. This is attributed to the
width of the band gap between the lowest two excited bands reaching a
minimum for certain value of V$_{1,}$ as is expected when considering that
the bandgap for this phase shift value is diminished by destructive
interference of the amplitudes of second order Bragg scattering of the
standing wave potential and first order Bragg scattering of the potential
with periodicity $\lambda /4$. The inset of Fig.~5 is to indicate the
dependence of the theoretical
value of the bandgap as a function of V$_{1}$. We interpret the
experimental data of Fig.~5 as clear evidence for the destructive
(constructive) interference of scattering amplitudes contributing to
the size of the bandgap at a phase shift of $\varphi=180^\circ$
($\varphi=0^\circ$) respectively between lattice harmonics.
\begin{figure}
\includegraphics[width=0.9\linewidth]{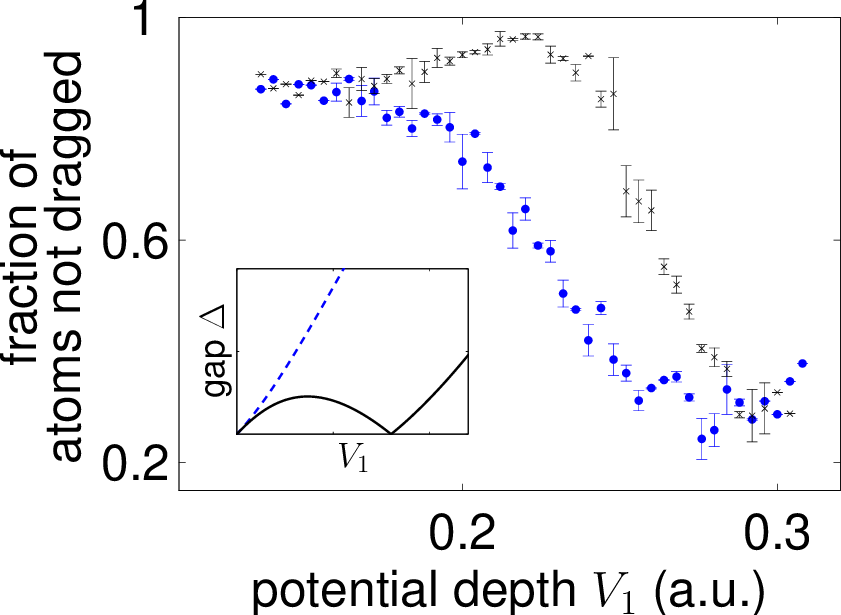}%
\caption{Fraction of atoms that have tunneled from the first to the
  second excited
  band as a function of the potential depth V$_{1}$ of the standing
  wave contribution to the total lattice potential for a phase
  $\varphi =0^\circ$ (dots) and $\varphi =180^\circ$ (crosses) between
  lattice harmonics. The inset indicates the variation of the expected
  gap size on V$_1$ for $\varphi=0^\circ$ (dashed line)
  and $\varphi=180^\circ$ (solid line).}
\end{figure}
To conclude, we have studied the band structure of optical lattices with
variable spatial symmetry and shape by means of quantum transport of an
atomic Bose-Einstein condensate. We find that the Landau-Zener tunneling
rate between the first and second excited Bloch band depends critically on
the phase between spatial Fourier components of the lattice, which is
attributed to interference effects within the band spectrum.

For the future, we expect that optical lattices of nonstandard shape allow
for novel quantum gas phases, and model solid state physics problems such as
quantum magnetism and frustrated
lattices~\cite{Damski-kagome,Lukin-spin,Santos-kagome}. A different perspective includes
quantum ratches with atomic
Bose-Einstein condensates~\cite{Flach-ratchet}. An exporation of the Hamiltonian ratchet
regime is expected to allow for novel quantum dynamical phenomena.

We thank P. H\"{a}nggi for discussion in which the idea to study
Landau-Zener tunneling in lattices of variable shape arose. We acknowledge
financial support of the Deutsche Forschungsgemeinschaft.

\small{


\clearpage


\begin{thebibliography}{10}

\bibitem{kittel}
See e.g.: C. Kittel, {\it Introduction to Solid State Physics}, (John
Wiley {\&} Sons, New York,1956).

\bibitem{Bloch-lattice}
See e.g.: I. Bloch, Nature Physics \textbf{1}, 23 (2005).

\bibitem{Salomon-Bloch}
M. Ben Dahan {\it et al.}, Phys. Rev. Lett. \textbf{76}, 004508 (1996)

\bibitem{Raizen-Wannier}
Q. Niu, X.-G. Zhao, G. A. Georgakis, and M. G. Raizen,
Phys. Rev. Lett. \textbf{76}, 4504 (1996)

\bibitem{Raizen-Landau}
C. F. Bharucha {\it et al.}, Phys. Rev. A \textbf{55}, R857 (1997)

\bibitem{Eisenstein-heterostructure}
J. P. Eisenstein {\it et al.}, Phys. Rev. Lett. \textbf{53}, 2579 (1984)

\bibitem{Renzoni-dissi}
R. Gommers, S. Bergamini, and F. Renzoni, Phys. Rev. Lett. \textbf{95},
073003 (2005)

\bibitem{Korsch-Zener}
B. M. Breid, D. Witthaut, and H. J. Korsch, New J. Phys. \textbf{9}, 62
(2007)

\bibitem{Bloch-Theory}
F. Bloch, Z. Phys. \textbf{52}, 555 (1928)

\bibitem{Ritt-Fourier}
G. Ritt {\it et al.}, Phys. Rev. A \textbf{74}, 063622 (2006)

\bibitem{Cohen-multi}
P. R. Berman, B. Dubetsky, and J. L. Cohen, Phys. Rev. A \textbf{58},
4801 (1998)

\bibitem{Weitz-multi}
M. Weitz, G. Cennini, G. Ritt, and C. Geckeler,
Phys. Rev. A \textbf{70}, 043414 (2004)

\bibitem{Cennini-BEC}
G. Cennini, G. Ritt, C. Geckeler, and M. Weitz,
Phys. Rev. Lett. \textbf{91}, 240408 (2003)

\bibitem{Mellish-ground}
A. S. Mellish, G. Duffy, C. McKenzie, R. Geursen, and A. C. Wilson,
Phys. Rev. A \textbf{68}, 051601(R) (2003)

\bibitem{Damski-kagome}
B. Damski {\it et al.}, Phys. Rev. Lett. \textbf{95}, 060403 (2005)

\bibitem{Lukin-spin}
L.-M. Duan, E. Demler, and M. D. Lukin, Phys. Rev. Lett. \textbf{91},
090402 (2003)

\bibitem{Santos-kagome}
L. Santos {\it et al.}, Phys. Rev. Lett. \textbf{93}, 030601 (2004)

\bibitem{Flach-ratchet}
S. Denisov, L. Morales-Molina, S. Flach, and P. H{\"a}nggi,
Phys. Rev. A \textbf{75}, 063424 (2007)

\end{thebibliography}
\end{document}